**Doubly Stabilized Perovskite Nanocrystal Luminescence Downconverters**

*Qi Xue\*, Carola Lampe\*, Tassilo Naujoks, Kilian Frank, Moritz Gramlich, Markus Schoger, Willem Vanderlinden, Patrick Reisbeck, Bert Nickel, Wolfgang Brütting, Alexander Urban\**

Dr. Qi Xue, Carola Lampe, Moritz Gramlich, Markus Schoger, Prof. Dr. Alexander Urban
Nanospectroscopy Group, Nano-Institute Munich, Faculty of Physics,
Ludwig-Maximilians-Universität München, 80539 Munich, Germany
E-mail: qi.xue@lmu.de, carola.lampe@physik.uni-muenchen.de, urban@lmu.de

Tassilo Naujoks, Prof. Dr. Wolfgang Bruetting
Organic Semiconductors Group, Institute of Physics,
University of Augsburg, 86159 Augsburg, Germany

Dr. Willem Vanderlinden
Chair of Applied Physics and Center for NanoScience,
Ludwig-Maximilians-Universität München, Amalienstr. 54, 80799 Munich, Germany

Kilian Frank, Patrick Reisbeck, PD Dr. Bert Nickel
Soft Condensed Matter Group, Faculty of Physics,
Ludwig-Maximilians-Universität München, 80539 Munich, Germany

## Keywords

## Abstract

Halide perovskite nanocrystals (NCs) have emerged as a promising material for applications ranging from light-emitting diodes (LEDs) to solar cells and photodetectors. Still, several issues impede the realization of the nanocrystals' full potential, most notably their susceptibility to degradation from environmental stress. This work demonstrates highly stable perovskite nanocrystals (NCs) with quantum yields as high as 95 % by exploiting a ligand-assisted copolymer nanoreactor-based synthesis. The organic ligands thereby serve a dual function by enhancing the uptake of precursors and passivating the NCs. The polymer micelles and ligands thus form a double protection system, shielding the encapsulated NCs from water-, heat- and UV-light-induced degradation. We demonstrate the optoelectronic integrability by incorporating the perovskite NCs as spectrally pure downconverters on top of a deep-blue-emitting organic LED. These results establish a way of stabilizing perovskite NCs for optoelectronics while retaining their excellent optical properties.

**Main Text**

Quantum Dots (QDs), semiconductor nanocrystals (NCs) exhibiting strong quantum-confinement, have incredibly appealing properties for display and lighting applications, including bright and color-pure photoluminescence (PL) and large spectral tuning ranges.[1, 2] Electroluminescence from a QD light-emitting device (LED) was first demonstrated in 1994, and modern architectures build on the more prevalent organic light-emitting diodes - OLEDs.[1, 3-5] With charge balancing and high efficiency across the whole visible spectrum still far from optimal, QDs are typically used as color-pure luminescence downconverters for displays in backlighting inorganic LEDs.[2, 6] While research into device architectures is critical for future development, many groups are also exploring alternative materials for QDs to reduce cost, minimize abundance concerns, and limit toxicity.[7-9] Having initially emerged in 2012 as a promising material for photovoltaics,[10] halide perovskites possess significant potential for nearly all optoelectronic devices, such as photodetectors,[11] lasers,[12] and LEDs.[13-15] Halide perovskite NCs are especially advantageous for light emission, with their emission wavelength tunable throughout the visible range by composition and morphology,[16] exceedingly high quantum yields (QYs), even approaching unity[17-19], and syntheses generally facile, cheap and easily scalable.[20, 21] In contrast, traditional QDs typically require inorganic core-shell structures, high precursor purity, and complex, tedious syntheses to achieve impressive optical properties.[22] However, perovskite

NCs lack sufficient stability to enable widespread commercialization.[23] Therefore, much research has concentrated on enhancing NC stability by embedding the NCs inside matrices,[24, 25] by optimizing ligands,[26] or by developing core-shell systems.[27, 28] Each strategy has advantages and limitations, such as reduced quantum yield (QY), stability, or limited charge injection. Recently, we developed a method employing block copolymers, which function both as nanoreactors and stabilizers for the perovskite NCs.[29] These individually wrapped NCs were remarkably stable, with thin films able to survive full underwater submersion for nearly 80 days. Nevertheless, PLQYs in dispersion did not surpass 63 %, and the overall yield of NCs was relatively low. This is likely because we did not employ any organic ligands, which are typically crucial for obtaining high-quality halide perovskite NCs.[30, 31] The ligands serve a dual purpose, enhancing the solubility of the perovskite precursors and passivating undercoordinated ions at the NC surfaces, minimizing non-radiative recombination. Here, we merge both synthetic strategies by adding organic ligands to the block copolymer-based synthesis to improve the efficiency of our already highly stable NCs. The resulting dispersions are highly luminescent, with QYs up to 95%. We confirm the inner structure of the NCs and their assemblies by combining transmission electron microscopy (TEM), X-ray diffraction (XRD), and linear optical spectroscopy. Interestingly, we find that the length of the organic ligand has a strong effect on the emission wavelength of the NCs, *i.e.,* shorter ligands induce stronger quantum confinement. Most importantly, the ligand plus polymer stabilized NCs are markedly less susceptible to degradation induced by exposure to water, UV-light, and heat. The enhanced stability allows for optoelectronic integration; we demonstrate their applicability as efficient luminescence downconverters on top of a deep-blue-emitting OLED. These results highlight that the doubly stabilized NCs present a substantial advancement for obtaining simultaneously stable and efficient perovskite NC emitters and could find widespread use in displays or even optically-pumped lasers.

For the synthesis, we dissolve a diblock copolymer comprising polystyrene and poly(2-vinyl pyridine) (PS-b-P2VP) in toluene.[29] Above the critical micelle concentration (CMC), the polymers spontaneously undergo phase separation and form uniform core-shell micelles with the P2VP forming the core surrounded by a PS shell. A methylammonium halide precursor (MAX, X = Br, I) is added to the solution under vigorous stirring to enable diffusion into the micellar cores, where the environment is more polar. The first part of this synthesis is identical to our previously published method. However, for these NCs here, we now add a solution comprising the lead halide precursor and organic ligands (aliphatic amines and acids). The dispersions show an immediate increase in coloration, which continually intensifies over time. This is a clear indication that this step substantially improves the perovskite NCs' formation process. To quantify this observation, we obtained linear photoluminescence (PL) spectra of the brightly colored dispersions (Figure 1a). All of these exhibited strong PL emission, characterized by a single, narrow peak (<120 meV full width at half maximum - FWHM). The emission maxima varied between 520 nm and 716 nm, depending on the bromide to iodide ratio and ligands used in the precursor. The PLQY of all dispersions – even the mixed halides – were exceedingly high, peaking at 95 % (for the pure bromide composition), far above our previously obtained maximum of 63 %.[29]

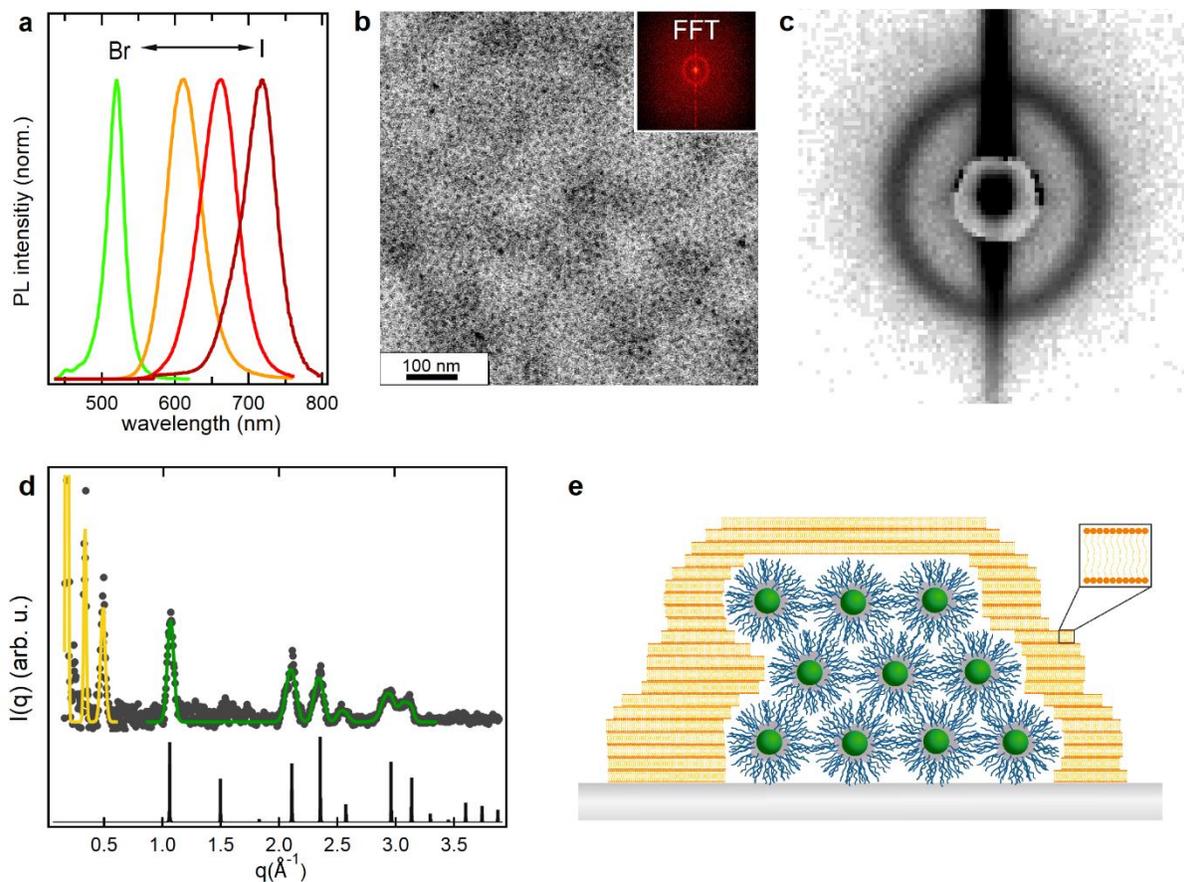

**Figure 1. Formation and morphology of perovskite NCs.** (a) PL spectra of doubly stabilized MAPbBr$_x$I$_{3-x}$ NCs with $0 \leq x \leq 3$. (b) TEM image of doubly stabilized MAPbBr$_3$ NCs. The NCs self-assemble to form a monolayer and exhibit a homogeneous core size of 8±2 nm. FFT of the image (inset) results in a value of 19±1 nm for the NC spacing. (c) GISAXS detector image of bare polymer micelles mixed with dodecylamine ligands revealing a spacing of 20.7 nm. (d) GIWAXS data of doubly stabilized MAPbBr$_3$ NCs with oleylamine ligands (grey dots) show Bragg peaks matching DFT calculations of the cubic MAPbBr$_3$ structure (black lines).[32] Scherrer analysis of the peak widths (green line) yields NC core sizes of 10.3±2.6 nm. Sharp peaks at small scattering angles (yellow line) correspond to a lamellar ligand bilayer phase with a 3.96 nm spacing. (e) Schematic of a NC film on substrates composed of tightly packed doubly stabilized NC assemblies embedded in an excess ligand phase.

Transmission electron microscopy (TEM) was employed to investigate the NCs' morphology (Figure 1b). The NCs exhibited a narrow size distribution with diameters of 8±2 nm and formed densely-packed self-assembled monolayers upon drop-casting on the TEM grids. A fast Fourier transform (FFT) of the TEM image reveals a center-to-center distance of approximately 19 nm (inset in Figure 1b). AFM imaging of the NC films confirms this self-assembly process and the obtained sizes (see Supporting Information Figure S1) and an excellent homogeneity of the sample. Analysis reveals nearly perfect monolayers, 16 nm in height, slightly above the value obtained from the TEM images, over hundreds of µm² with a typical surface roughness of 1.5 nm.

We used X-ray scattering to investigate the morphology and ordering of the doubly stabilized NCs in films. Grazing-incidence small-angle X-ray scattering (GISAXS) data of such dried multilayer films on a silicon substrate are shown in Figure 1c and Figure S2. The GISAXS data reveal a circular diffraction signal around the direct beam at $q = \frac{4\pi}{\lambda} sin\left(\frac{2\theta}{2}\right) = 0.035 \text{ Å}^{-1}$. The homogeneous powder-ring-like diffraction image indicates an isotropic packing of the micelles in all three dimensions. We obtain a center-to-center distance of the micelles of $d = 20.7 \ nm$ assuming hexagonal close-packing (see Table S1). This size nearly equals the value obtained from TEM analysis. To investigate the perovskite crystallinity, we also recorded grazing-incidence wide-angle scattering (GIWAXS) of the NCs (Figure 1d and Figure S2). We find pronounced diffraction peaks (green line) whose positions reveal a near-perfect match to those expected from a cubic MAPbBr$_3$ structure (black bars).[32] A Scherrer analysis of the MAPbBr$_3$ peak widths reveals a NC size of (10.3±2.6 nm), matching the core sizes observed in the TEM images. Interestingly, at small scattering angles ($q < 0.5 \text{ Å}^{-1}$), we find a pronounced additional diffraction signal stemming from a lamellar structure ordered parallel to the substrate (yellow line, see also Figures S2 and S3). Analysis of these peaks reveal a lamellar structure with a spacing of 3.96±0.03 nm. Being approximately twice the length of an oleylamine molecule,[33] we attribute this signal to a multilamellar ligand phase comprising the oleylamine and oleic acid molecules and domain sizes of at least 30 nm.[33] Accordingly, we envision the films to consist of MAPbBr$_3$ NCs passivated with organic ligands and incorporated inside the polymer micelles. These doubly stabilized NCs self-assemble into densely-packed layers, with residual ligands forming lamellar bilayers, oriented parallel to the substrate and surrounding the micellar-encapsulated NCs.

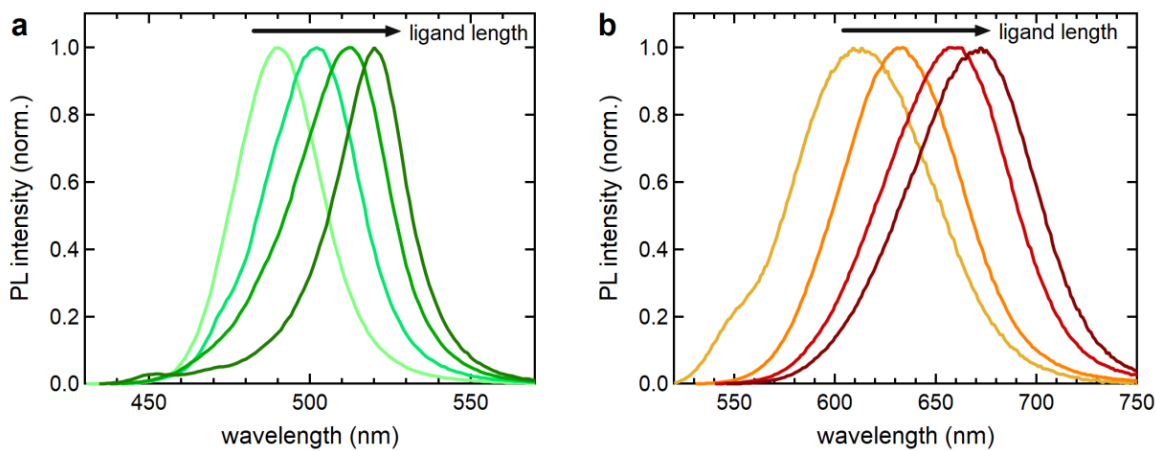

**Figure 2. Effect of organic ligand length on NC properties.** (a) PL spectra of MAPbBr$_3$ NCs synthesized with varying amine ligand lengths (from left to right hexylamine, octylamine, dodecylamine, oleylamine). (b) PL spectra of MAPbI$_3$ NCs synthesized with varying amine ligand lengths (from left to right hexylamine, octylamine, dodecylamine, oleylamine).

The GIWAXS data suggested that the encapsulated NCs are embedded in a lamellar bilayer structure formed by excess ligands. Therefore, we varied the ligand length to probe whether they influence the synthesis beyond enhancing the precursors' solubility. Indeed, in contrast to the precursor concentration, the amine ligand's length had a pronounced effect on the spectral position of the PL emission maximum (Figure 2). The PL emission maximum of MAPbBr$_3$ NCs blueshifts from 520 nm down to 491 nm as we shorten the alkyl chain from 18 carbon atoms (oleylamine) down to 6 (hexylamine). UV-vis spectra also showed a progressive blueshift of the absorption onset from 520 nm to 491 nm with decreasing chain length for oleylamine and hexylamine, respectively (see Figure S7a). For MAPbI$_3$ NCs, we observe a similar behavior with a shift from 670 nm to 610 nm of the PL emission maximum (see Figure 2b). This blueshift suggests that the ligands induce quantum confinement in the NCs. To confirm this, we imaged the NCs again, utilizing TEM. The NCs all form homogeneous monolayers on the TEM grids. Astoundingly, the NC core size and center-to-center distance exhibit negligible dependence on the amine ligand length for all samples (see Figure S8). To explain these results, we must assume that each micelle contains more than one NC on average, each small enough to induce quantum confinement and unresolvable in the TEM images. Compared to previous reports on quantum-confined NCs, we surmise that the NCs responsible for the PL here are between 3 nm and 15 nm in size.[21, 34, 35] If the NCs do not fill the entire core space, there will likely be a more considerable inhomogeneity in the NC sizes, increasing the PL peak width. This is supported by the PL peak widths, which are considerably broader for the shorter ligand-containing samples than those synthesized with oleylamine (170 meV compared to 118 meV).

Hexylamine is the shortest viable amine ligand; even shorter ones, e.g., butylamine, resulted in very low PL emission (see Figure S9). The high volatility of the shorter amines, which causes them to evaporate from the solution, likely reduces their concentration significantly, prohibiting them from participating during the synthesis. Notably, the PLQY assumed high values for all viable ligands, with a maximum average value of 90% for octylamine and a minimum of 78% for dodecylamine. Interestingly, the acidic ligand plays a less significant role, i.e., varying the length of the acid chain from 18 to 6 C-atoms shifted the PL spectra by only 9 nm, in contrast to the 29 nm shift obtained for the amine chain length variation (see Figure 2 and Figure S10). Moreover, PLQYs were generally significantly reduced for the other acid ligands.

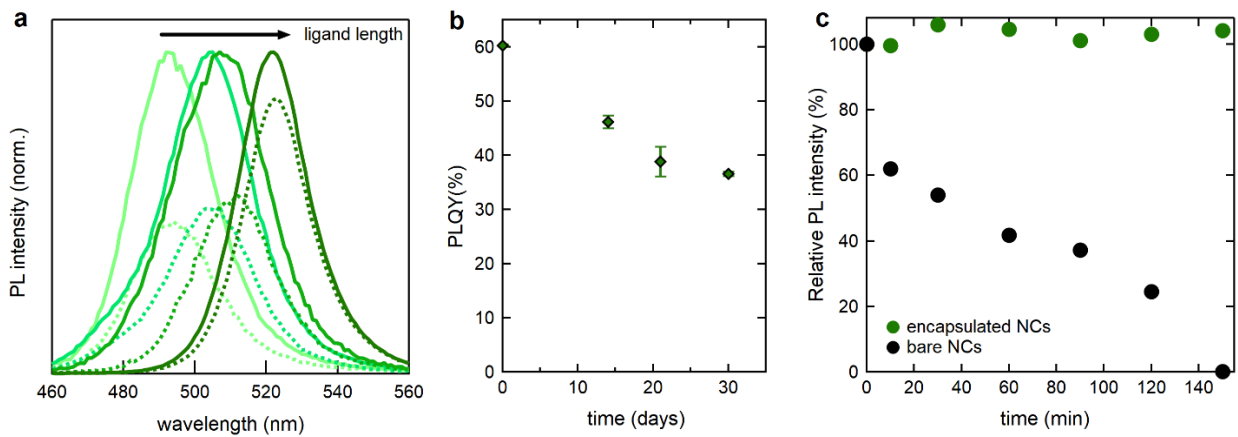

**Figure 3. Stability of doubly stabilized perovskite NCs.** (a) PL spectra of dropcasted doubly stabilized perovskite NC films with varying amine ligand length immediately after synthesis (solid lines) and after submersion underwater for 12 hours (dashed lines). (b) Evolution of the PLQY of doubly stabilized perovskite NC films stored in ambient conditions for 30 days. (Three films were fabricated, and their average PLQY and the standard deviation determined). (c) Development of the PL intensity of perovskite NC dispersions for doubly stabilized NCs (green) and ligand encapsulated NCs (black) illuminated with UV light.

In our previous publication, we found that the polymer micelle encapsulation greatly enhanced the stability of the perovskite NCs towards environmentally-induced degradation, even permitting films to be stored underwater for up to 75 days.[29] The degradation progressed much faster initially and then slowed down considerably. We repeated similar measurements to determine whether the organic ligands help deter degradation, producing drop-casted films of the doubly protected perovskite NCs. To minimize the protective effect of thick drop-casted films, we fabricated films with as little material as possible to obtain reliable PL signals. We submersed these in water, monitoring the PL spectra and intensities over the first 12 hours (Figure 3a). All films remained luminescent, exhibiting PL spectra with nearly no change in the shape, only lower overall intensities. Interestingly, the intensity decrease was highly dependent on the ligand length used. While the hexylamine sample's PL strength decreased to 46%, the oleylamine-protected NCs retained 86% of their original PL signal. This effect is likely due to the higher hydrophobicity of the longer chains, which shield the NCs more effectively. The NCs are not only bright but also highly efficient, with films comprising oleylamine doubly protected NCs exhibiting quantum yields of 60% (Figure 3b). The QY of these otherwise unprotected films shows a gradual yet slow decrease to 37% after exposure to ambient conditions for 30 days. This constitutes a significant improvement over most nanocrystalline or thin perovskite films.[26, 36, 37] Additionally, the films are relatively stable to heat and UV light exposure, showing little change in the PL spectra or overall intensity (Figure 3c and Figure S11). We investigated the stability regarding UV exposure in dispersions to exclude the combined effect of oxygen and UV exposure (Figure 3c). Reference $MAPbBr_3$ NCs synthesized through the ligand-assisted reprecipitation (LARP) method without the block copolymer rapidly degrade when exposed to 365 nm light, with the PL intensity dropping to zero after only 150 minutes. In comparison, the doubly stabilized NCs did not exhibit a discernible drop in their PL even for significantly longer exposure times (up to 24 hours). This stability is especially relevant for device fabrication, as the NCs do not need to be processed immediately and instead can be stored in ambient conditions until required.

The doubly stabilized NCs are substantially more stable than typical perovskite NCs; however, the additional organic material in the structure might impede integration into optoelectronic devices. To investigate this, we employed the NCs as efficient luminescence downconverters with highly pure emission, similar to those used in QD-TVs.

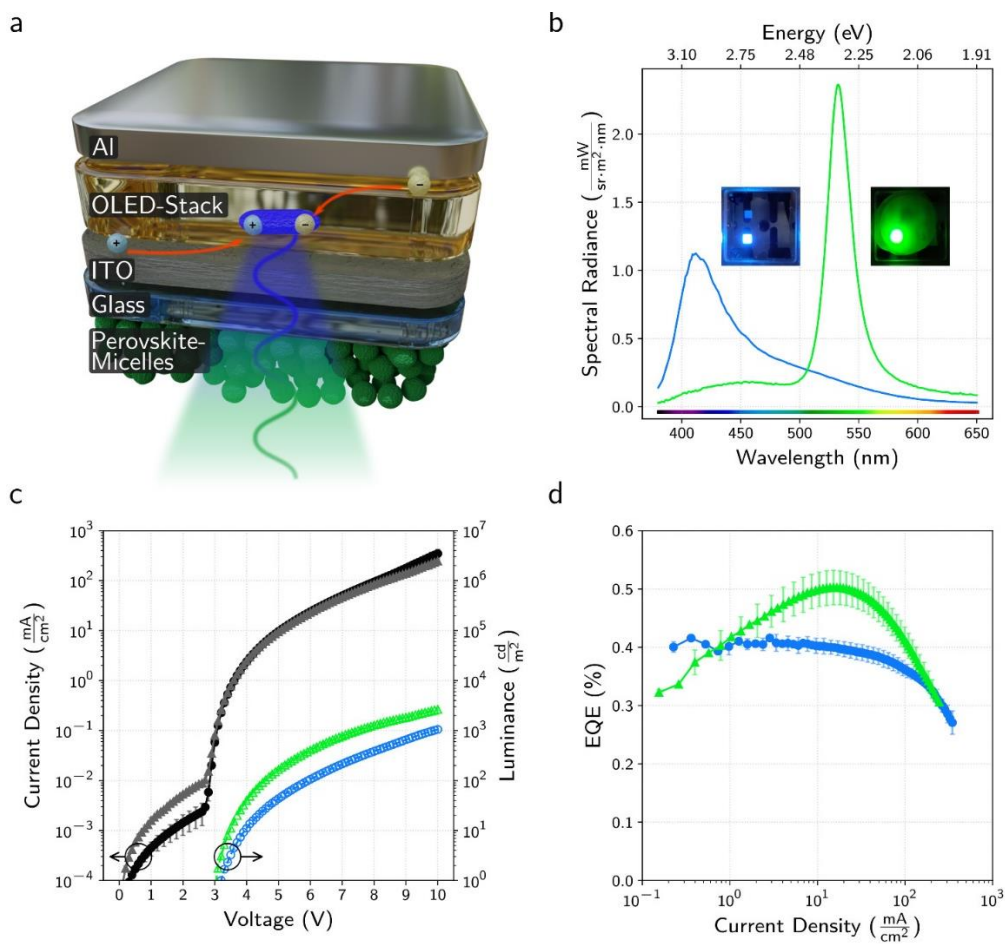

**Figure 4. Perovskite NC downconverter OLED.** (a) Schematic layer stack of the blue pump OLED on the top side of a glass substrate with the perovskite NC downconversion layer applied to the bottom side of the substrate. (b) Electroluminescence (EL) emission of the bare OLED (blue) and the complete device (green). Photographs of the two devices are shown as insets. (c) Current density and luminance of the bare OLED (black, blue) and the entire device (grey, green) dependent on the driving voltage. (d) External quantum efficiency (EQE) of the bare OLED (blue) and the DC-OLED (green) as a function of the current density.

To this end, we fabricated an OLED comprising a deep-blue fluorescent emitter (Figure 4; for details on the layer stack and the used materials, see the SI: Figures S12 and S13). This OLED has a low turn-on voltage of about 3 V, a spectral emission peaking at 405 nm, and reaches a luminance of more than 1000 cd m$^{-2}$ at 10 V. The OLED's external quantum efficiency (EQE) is about 0.4 % at low current density before it starts to roll off above 10 mA cm$^{-2}$, which can be ascribed to exciton-exciton and exciton-polaron quenching.[38] We fabricated a green downconverter OLED (DC-OLED) by drop-casting a layer of highly concentrated doubly protected NCs comprising oleylamine and oleic acid ligands onto the back of the deep-blue OLED's glass substrate, as shown schematically in Figure 4a. The perovskite layer absorbs nearly all blue light emanating from the OLED pump and converts it into green light. The resulting device is strongly luminescent with an electroluminescence emission peaking at 532 nm and a very narrow FWHM of only 24 nm (Figure 4b). As expected, the DC-OLED shows almost identical electrical characteristics; however, the luminance reaches substantially higher values than the OLED for all driving voltages and achieves a maximum of 2650 cd m$^{-2}$ at 10 V, which was the highest voltage used (Figure 4c). This increase in luminance is understandable as the human eye is more sensitive toward the green part of the visible spectrum. Most remarkable, however, is the enhancement of the external quantum efficiency (EQE) of the DC-OLED (Figure 4d).

In detail, the DC-OLED EQE, initially below that of the original OLED, increases steeply with increasing current density, rapidly surpassing the formers' performance and reaching a peak value of about 0.5% at 20 mA/cm² (corresponding to a luminance of about 350 cd/m²) and then decreasing rapidly. Such behavior is frequently observed in quantum-dot LEDs because the total recombination current has three contributions, scaling with different power of the carrier density: a linear (non-radiative) term owing to trap-assisted recombination, a quadratic term for the radiative recombination, and, finally, a cubic term for Auger quenching.[39, 40] Since the EQE is defined as the ratio of the radiative and the total recombination currents, it typically has a maximum at intermediate currents, where the quadratic term is the dominant contribution. Following this scenario for the DC-OLEDs studied here, the "roll-on" behavior can tentatively be ascribed to trap filling in the perovskites with increasing photon flux from the blue OLED pump. A concomitant increase in PLQY with increasing excitation fluence, occurring in the neat DC film, supports this explanation (see the Experimental Section for details). On the other hand, the steep "roll-off" above the EQE maximum is a superposition of the droop of the OLED itself, plus an additional contribution stemming from Auger recombination in the perovskite micelles.[41] Further investigations toward clarifying the precise nature of these processes are underway.

The enhancement of the EQE over a wide range of current densities is remarkable. The higher EQE is a priori not expected if only a simple luminescence downconversion process is assumed, where each absorbed blue pump photon can, at best, produce one green photon in the perovskite. It is even more surprising given the less than ideal PLQY of the doubly stabilized perovskite film of about 71% (see Fig. S15), which should lead to a decrease in the EQE. Consequentially, this indicates that the perovskite layer on top of the glass substrate itself improves light outcoupling from the blue OLED pump. This, in turn, overcompensates the losses in the luminescence downconversion process. Although exact quantification of the overall enhancement factor is not straightforward (see the Experimental Section for details), one can qualitatively understand this boost by considering the effective refractive index of the perovskite-polymer composite layer. As this is higher than glass, significantly more blue photons – about a factor of 2 – from the OLED pump are extracted into this layer compared to the glass-air interface in the pristine device. Moreover, aggregates of encapsulated perovskite NCs in the applied layer scatter a considerable fraction of these green photons out of the device.

In summary, we have synthesized perovskite NCs inside block copolymer nanoreactors with photoluminescence quantum yields of up to 95 % by incorporating organic ligands into the synthesis. The organic ligands help solubilize the perovskite precursors, promoting their diffusion into the micellar nanoreactor cores. Therein, the perovskite NCs form spontaneously and are efficiently stabilized by both the diblock copolymer micelle and the organic ligands. This leads to enhanced quantum confinement and, consequently, a tuning of the PL emission. The polymer and organic ligands form a stabilized system around the NCs, shielding them from the environment and substantially reducing degradation due to water, heat, and UV-light exposure. Since the double stabilization system is optically transparent, integration into optoelectronic devices is straightforward. We demonstrate this by fabricating an efficient downconverter LED with strong, spectrally narrow green luminescence. The novel structured NCs exhibit a respectable combination of stability and optical properties and could pave the way to commercialize perovskite NC-based optoelectronics. Initially used as luminescence downconverters for displays, after resolving charge and energy transport issues, doubly stabilized perovskite NCs could represent an alternative material for realizing directly electrically contacted QD-LEDs.

**Experimental Section**

*Chemicals*

Methylammonium bromide (MABr, 99.5%) and Methylammonium iodide (MAI, 99.9%) were purchased from Ossila. Lead bromide (PbBr$_2$, 98%), butylamine (99%), hexylamine (99%), octylamine (99%), dodecylamine (99%) oleic acid (technical grade 90 %) and oleylamine (technical grade 70 %,), were purchased from Sigma-Aldrich. Lead iodide (99.9%) was purchased from Alfa Aesaer. Toluene (99.8%) was purchased from VWR chemicals. Block Copolymers were purchased from Polymer Source Inc. All chemicals were used without further purification.

*Doubly stabilized perovskite nanocrystal synthesis*

Briefly, the perovskite NCs were prepared through a wet chemical synthesis method. To prepare the polymer micelle solution, a PS-PVP polymer ($9.536 \cdot 10^{-7}\ mol$) was dissolved in toluene (3 ml) and continuously stirred overnight. Then, MAX ($2.14 \cdot 10^{-5}\ mol$; X= Br, I) was added into the polymer solution, followed by vigorous stirring for 5 hours. A lead halide salt ($0.1\ mmol$) was dissolved in 0.5 ml toluene (500 µl) with an acid ligand ($0.14\ mmol$) and an amine ligand ($0.38\ mmol$) as the lead precursor. The lead precursor (30 µl) was added into the MAX-polymer mixture and stirred for at least 24 hours. The reaction mixture was centrifuged for 10 min at 10000 rpm for further purification. The supernatant was subsequently centrifuged for 30 min at 12000 rpm and further used for characterization.

*$MAPbBr_3$ NCs synthesized through the ligand-assisted reprecipitation (LARP) method*

Typically, MABr ($0.16\ mmol$) and $PbBr_2$ ($0.2\ mmol$) were added into dimethylformamide (DMF) (5 ml), followed by n-octylamine (20 µl) and oleic acid (0.5 ml) to obtain a precursor solution. Under vigorous stirring, precursor solution (2 ml) was added into toluene (10 ml). The reaction mixture was centrifuged for 10 min at 7000 rpm to obtain a greenly fluorescing supernatant with NCs.

*Transmission electron microscopy (TEM)*

The morphology of the perovskite NCs was characterized by transmission electron microscopy (TEM) operating at an accelerating voltage of 80 kV (JEOL JEM-1011).

*Atomic force microscopy (AFM)*

Atomic force microscopy imaging was performed on a JPK Nanowizard Ultrapseed 2 AFM (Bruker). Images were recorded in amplitude-modulation mode on dried samples, under ambient conditions, and using silicon cantilevers (Olympus; AC160TS; resonance frequency ~250- 300 kHz). Typical scans were recorded at 1-3 Hz line frequency, with optimized feedback parameters, and at ~2nm/pixel. The Scanning Probe Image processor software (SPIP;v6.4; Image Metrology) was employed for image processing and analysis. Image processing involved background correction using a global fitting with a third-order polynomial and line-by-line correction through the histogram alignment routine.

*Grazing-incidence small/wide-angle scattering (GISAXS/GIWAXS)*

(GI)SAXS and (GI)WAXS data were recorded at the molybdenum microfocus X-ray setup at LMU Munich.[42] The X-ray energy is 17.4 keV (wavelength λ = 0.71 Å). Grazing incidence small- and wide-angle X-ray scattering (GISAXS and GIWAXS) were recorded from samples dried on 1x1 cm² silicon substrates (*Siegert Wafer*) at a grazing incidence angle of 0.2°.

For GISAXS, either a Pilatus 300K (Dectris, Baden-Dättwil, Switzerland) detector at 2.53 m distance or a Pilatus 100K detector at 1.05 m distance was used. The total exposure time per sample was between 3 and 5 h.

For GIWAXS, a Pilatus 100K detector at 23.5 cm distance was raster-scanned perpendicular to the direct beam to increase the *q*-range. The total exposure time per sample was between 10 and 15 h.

We used the median of multiple detector images for background reduction for further processing. The resulting images were processed with the Nika package[43] for Igor Pro (Wavemetrics, Portland, OR, USA). For GISAXS, the beam stop's shadow, scattering around the specular axis $q_{//}$ = 0, and data below the sample horizon were masked before azimuthal integration. For GIWAXS, the beam stop's shadow, diffuse scattering from the silicon substrate, and data below the sample horizon were masked before azimuthal integration. A baseline was subtracted from the data in Origin Pro (OriginLab, Northampton MA, USA).

Data are shown as a function of scattering vector q:

$$q = \frac{4\pi}{\lambda} \sin\left(\frac{2\theta}{2}\right)$$

λ = 0.71 Å is the X-ray wavelength, 2θ is the scattering angle. In the grazing incidence geometry, we also show data as a function of the scattering vector in the substrate plane ($q_x$, $q_y$, $q_{||}$) and perpendicular to the substrate ($q_z$):

$$q_x = \frac{2\pi}{\lambda}(\cos(\alpha_f)\cos(\phi) - \cos(\alpha_i))$$

$$q_y = \frac{2\pi}{\lambda}\cos(\alpha_f)\sin(\phi)$$

$$q_\parallel = \sqrt{q_x^2 + q_y^2}$$

$$q_z = \frac{2\pi}{\lambda}(\sin(\alpha_i) + \sin(\alpha_f))$$

$\alpha_i$ and $\alpha_f$ are the incidence and exit angles, respectively. We calculate the exit angles $\alpha_f$ and $\phi$ from the position of the direct beam $(x,y)$ on the detector.

$$\alpha_f = \arctan\left(\frac{y}{SDD}\right) - \alpha_i$$

$$\phi = \arctan\left(\frac{x}{SDD}\right)$$

$SDD$ is the sample-to-detector distance.

Peak widths were analyzed using Python and the *numpy*, *scipy*, *matplotlib*, *lmfit*, and *fabIO* packages.[44-47] We assume Gaussian peak broadening of a peak at $2\theta_1$ to a full width at half maximum (fwhm) $b_1$ due to the finite size of the nanocrystals. Additionally, there is an instrumental peak broadening. The main contribution to instrumental peak broadening is the sample length *l* =10 mm, which is approximately described by a boxcar peak shape between scattering angles $2\theta_0$ and $2\theta_2$. This boxcar function must be convoluted with a Gaussian peak of width $b_3$, which includes the broadening due to finite crystallite size and remaining instrumental contributions. This leads to a peak shape function $B(2\theta)$:

$$B(2\theta) = \frac{h}{N}\left[\text{erf}\left(\frac{2\theta_0 - 2\theta}{\sqrt{2}\sigma}\right) - \text{erf}\left(\frac{2\theta_2 - 2\theta}{\sqrt{2}\sigma}\right)\right]$$

$$N = \text{erf}\left(\frac{2\theta_0 - 2\theta_1}{\sqrt{2}\sigma}\right) - \text{erf}\left(\frac{2\theta_2 - 2\theta_1}{\sqrt{2}\sigma}\right)$$

$$2\theta_0 = \arctan\left(\frac{SDD \cdot \sin(2\theta_1)}{SDD \cdot \cos(2\theta_1) - l/2}\right)$$

$$2\theta_2 = \arctan\left(\frac{SDD \cdot \sin(2\theta_1)}{SDD \cdot \cos(2\theta_1) + l/2}\right)$$

$$\sigma = \frac{b_3}{2\sqrt{2\ln(2)}}$$

$h$ is the height of the peak. $N$ is a normalization factor. $\text{erf}(x) = \frac{2}{\sqrt{\pi}}\int_0^x e^{-\tau^2}d\tau$ is the error function. $2\theta$ is the scattering angle, $b_3$ is the fwhm of the Gaussian contribution to the peak. SDD is the sample-to-detector distance, $l$ is the sample length.

All remaining contributions to instrumental peak broadening (direct beam profile, detector pixel size, and X-ray wavelength spread) can be summarised in a Gaussian function of angle-dependent width $b_2$. We determine this width by fitting the diffraction peaks of a lanthanum hexaboride standard (NIST SRM 660c), taking into account its different sample length $l$ =1 mm in the function $B(2\theta)$.

The final peak width (fwhm) after correction of all instrumental contributions is then

$$b_1 = \sqrt{b_3^2 - b_2^2}$$

The crystallite size can be calculated by applying the Scherrer equation to the diffraction peak width.[48] We use the expression

$$D = \frac{K\lambda}{b_1 \cos(2\theta_1/2)}$$

$K = 1$ is a shape factor, $\lambda$ the X-ray wavelength, $b_1$ and $2\theta_1$ the peak width and diffraction angle as described above. We then report the mean and standard deviation of $D$ values obtained from fitting six diffraction peaks, as shown in Figures 1d and S3.

For comparison to existing crystallographic data, we simulated peak positions using the software Mercury 4.2.0.[49]

*OLED Device Fabrication*

Devices were fabricated with a structure of Borosilicate glass (0.7 mm) / ITO (90 nm) / PEDOT:PSS (CH8000, 40 nm) / MoO$_3$ (10 nm) / mCBP (5 nm) / TCTA (40 nm) / Cz-PS:TCTA (1:9 vol. ratio doping, 15 nm) / TPBi (45 nm) / LiF (0.5 nm) / Aluminium (100 nm), according to Figure S12, with the organic materials' structure and full names in Figure S13. The ITO-glass substrates were obtained from Thin Film Devices Inc. PEDOT:PSS CH8000 alias "CLEVIOS™ P VP CH 8000" from Heraeus Epurio is used as a smoothing polymer. The molybdenum oxide was bought from Sigma Aldrich and stored in N$_2$-atmosphere. So were all organic semiconductors, which were purchased from Luminescence Technology Corporation. After a cleaning procedure with UV grade acetone and isopropanol, the surface was treated in a UV-Ozone-Cleaner (Novascan Model PSD-UV4). PEDOT:PSS of type CH8000 was spin-coated at 3000 rpm and baked at 150 °C for 15 minutes in air. All other OLED materials were deposited by thermal evaporation at a pressure of less than $2 \cdot 10^{-4}$ Pa at deposition rates of 100 pm/s.

*OLED and DC-OLED operation*

The electrical measurements were performed with an Agilent E5263A source-measure unit. Luminance was determined with a calibrated photopic filter and photodiode combination by Gigahertz-Optik GmbH. The electroluminescent spectra were taken with a calibrated spectrometer by Jena Technische Instrumente GmbH "JETI specbos 1201 focus". The external quantum efficiency (EQE) was estimated with the acquired spectra, the applied current and a Lambertian emission profile approximation. After full characterisation of the pristine OLED, the downconverter layer was applied: two times 100 µl of the polystyrene enriched perovskite micelles solution were dropped onto the backside of the OLEDs' glass substrate and dried at room temperature in nitrogen atmosphere resulting in a film of about a millimeter thickness. Intensity dependent photoluminescence quantum yield (PLQY) measurements were performed within a calibrated BaSO$_4$-coated integrating sphere via the three-step method reported previously. Excitation intensity was varied by using a continuous optical-density filter wheel in front of a 442 nm laser source. Detection of both the laser excitation and the PL signal was performed with a fiber-coupled CCD spectrometer (Princeton Instruments PyLoN).

*Estimating down-conversion efficiency and light out-coupling efficiency*

As discussed in the main manuscript, the DC-OLED shows a higher external quantum efficiency than the original pump OLED. This is a priori unexpected because in a simple luminescence down-conversion (color conversion) from blue to green photon multiplication is not possible. However, since a major fraction of photons is originally trapped in the blue pump OLED within waveguided and substrate modes, the quantum efficiency can be boosted by releasing these. It has been shown that high index out-coupling structures can lead to significant EQE increases.[50] We propose that the perovskite-micelle film, which almost certainly has a refractive index greater than that of the used glass substrate, suppresses part of the total internal reflection at the glass-air interface. Moreover, due to its morphology it is also expected to scatter light quite efficiently and, therefore, effectively acts as a high index out-coupling medium.

The luminescence down-conversion efficiency (DCE) can be estimated form the ratio of the EQEs of the original blue pump OLED and the green DC-OLED. At the point where the spectra (see Figure S14 are taken, i.e. at a current density of about 25 mA/cm$^2$, these values are 0.39 % resp. 0.50 %, which leads to a DCE of 126 %. Furthermore, with the known PLQY, we can estimate the extent of the different involved factors: In the absence of any extra light outcoupling from the pump OLED, the DCE should be equal to the PLQY, i.e. 71% (see Figure S15), at best. However, if there is additional light outcoupling from the blue OLED into the perovskite and scattering of the converted green light out of this layer, which is in total described by a factor $n_{amb}$ (for ambient outcoupling), the DCE larger than 1 can be explained:

$$DCE = PLQY \cdot n_{amb}$$

Thus, the light out-coupling from the original blue OLED to the environment is increased by $n_{amb} = 177\%$ upon application of the doubly protected perovskite NC downconverter film.

It is further insightful to compare the emission spectra of both OLEDs, i.e. the original blue pump and the green DC-OLED. The emission spectrum of the DC-OLED is described as a superposition of the photoluminescence spectrum PL(λ) of the micelles and the driving OLED's electroluminescence spectrum EL(λ), weighted with the absorption spectrum A(λ) of the micelles, as shown in Figure S14 This superposition fits the EL spectrum of the DC-OLED quite well proving that absorption and subsequent re-emission by the micelles is indeed the main conversion process in that device. In detail, the following expression was used for the fit:

$$a \cdot EL(\lambda) \cdot (1 - s \cdot A(\lambda)) + b \cdot PL(\lambda - \lambda_0)$$

where a, b and s are adjustable parameters, and the wavelength offset $\lambda_0$ was fitted to be 7.4 nm. This offset can be ascribed to reabsorption in the thick perovskite-micelle film, which is different under EL conditions where excitation occurs by the blue OLED through the glass substrate, whereas PL excitation of a film is performed from the surface to air. Since the Stokes shift in perovskite nanocrystals is very small, such reabsorption is stronger for the high-energy side of the emission spectrum and, therefore, effectively leads to a red-shift for the EL spectrum.

Finally, the PLQY of a neat perovskite micelles film (i.e. without the blue pump OLED) was investigated in dependence of the excitation fluence (Figure S15). Although the 442nm laser excitation wavelength does not exactly match the maximum of the blue pump OLED's emission spectrum, it allows for a comparison with the current density dependent EQE of the DC-OLED shown in Fig. 4d of the main manuscript. Specifically, the lowest laser excitation fluence of $10^{-1}$ W/m² approximately corresponds to a current density of 0.1 mA/cm², i.e. well below the EQE maximum. With increasing laser intensity, the PLQY also increases by about 20% before saturating at a value of ca. 71% at higher intensities. This behaviour is often observed in luminescent quantum dots and is usually ascribed to trap filling.[40, 51] Consequently, the "roll-on" in the device EQE of the DC-OLED can most likely be attributed to the PLQY increase if the difference in excitation wavelength is neglected.

## Supporting Information

Supporting Information is available from the Wiley Online Library or from the author.

## Acknowledgments

We gratefully acknowledge support from the Bavarian State Ministry of Science, Research and Arts through the grant "Solar Technologies go Hybrid (SolTech)" and from the Deutsche Forschungsgemeinsschaft (DFG) under Germany's Excellence Strategy EXC 2089/1-390776260. T.N. and W.B. acknowledge financial support by the DFG within the Priority Program 2196 "Perovskite Semiconductors" (Project No. 424708673). This work was also supported by the European Research Council Horizon 2020 through the ERC Grant Agreement PINNACLE (759744), by the German Ministry for Education and Research (BMBF) through the projects "ELQ-LED" (13N14422) and "Lucent" (05K19WMA) and by the Center for NanoScience (CeNS) through a joint research project. Furthermore, we thank Matías Herran and Stefan Maier for aiding in the absorption measurements of our samples and Andreas Singldinger for graphics support. Q. Xue and C. Lampe contributed equally to this work.

**Table of Contents Entry**

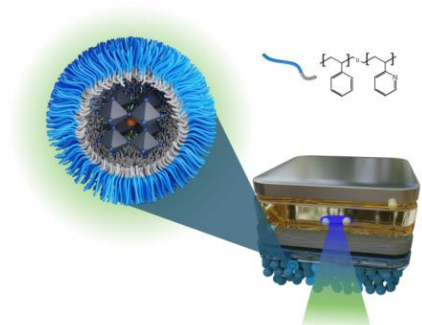

We present highly luminescent perovskite nanocrystals synthesized within a double stabilized system comprising block copolymers micelles and organic ligands. The double stabilization protect the NCs against environmentally-induced degradation and enhances the luminescence yielding quantum yields of up to 95%. The NCs are employed as optical downconverters on top of a deep blue OLED, providing bright, spectrally pure electroluminescence.

Supporting Information for

**Doubly Stabilized Perovskite Nanocrystal Luminescence Downconverters**

*Qi Xue\*, Carola Lampe\*, Tassilo Naujoks, Kilian Frank, Moritz Gramlich, Markus Schoger, Willem Vanderlinden, Patrick Reisbeck, Bert Nickel, Wolfgang Brütting, Alexander Urban\**

Dr. Qi Xue, Carola Lampe, Moritz Gramlich, Markus Schoger, Prof. Dr. Alexander Urban
Nanospectroscopy Group, Nano-Institute Munich, Department of Physics,
Ludwig-Maximilians-Universität München, 80539 Munich, Germany
E-mail: qi.xue@lmu.de, carola.lampe@physik.uni-muenchen.de, urban@lmu.de

Tassilo Naujoks, Prof. Dr. Wolfgang Bruetting
Organic Semiconductors Group, Institute of Physics,
University of Augsburg, 86159 Augsburg, Germany

Dr. Willem Vanderlinden
Chair of Applied Physics and Center for NanoScience,
Ludwig-Maximilians-Universität München, Amalienstr. 54, 80799 Munich, Germany

Kilian Frank, Patrick Reisbeck, PD Dr. Bert Nickel
Soft Condensed Matter Group, Department of Physics,
Ludwig-Maximilians-Universität München, 80539 Munich, Germany

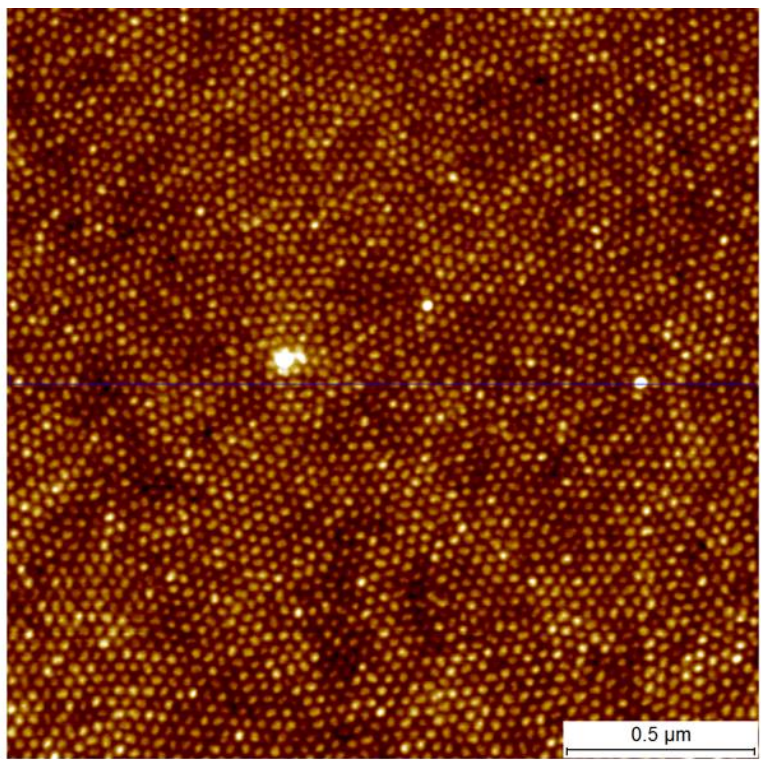
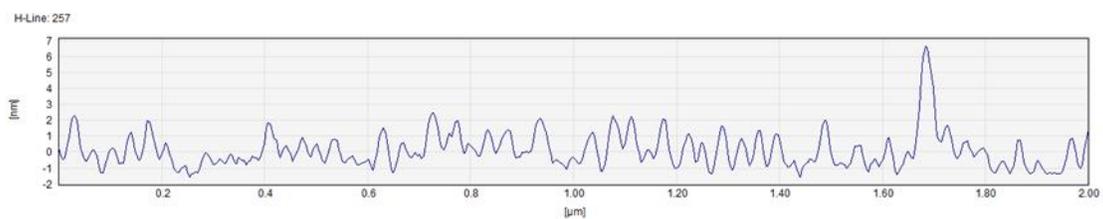

**Figure S1.** AFM image of double stabilized perovskite NCs with oleylamine and oleic acid as ligands. Onset is the corresponding height profile.

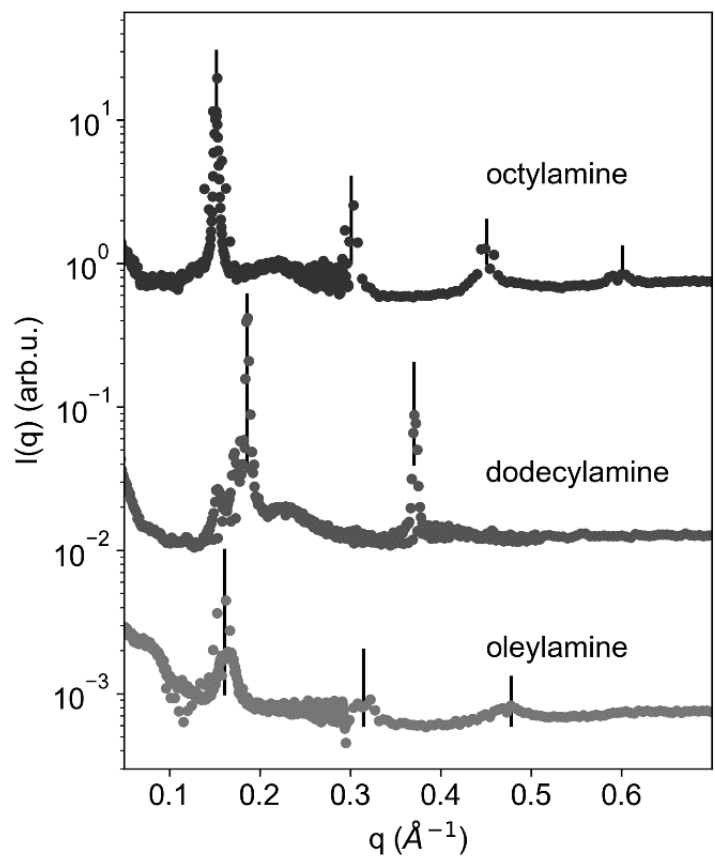

**Figure S2.** Azimuthally integrated GISAXS and GIWAXS data of dried samples (superimposed). Vertical lines indicate the diffraction peaks of lamellar phase of ligands. The lamellar spacings are (4.19 +/- 0.02) nm, (3.38 +/- 0.01) nm, and (3.96 +/- 0.03) nm for octylamine, dodecylamine, and oleylamine, respectively. The peak widths are limited by the instrumental resolution in the case of octylamine and dodecylamine (domain size > ca. 80 nm). In the case of oleylamine, the Scherrer equation yields a domain size of (30 +/- 11) nm.

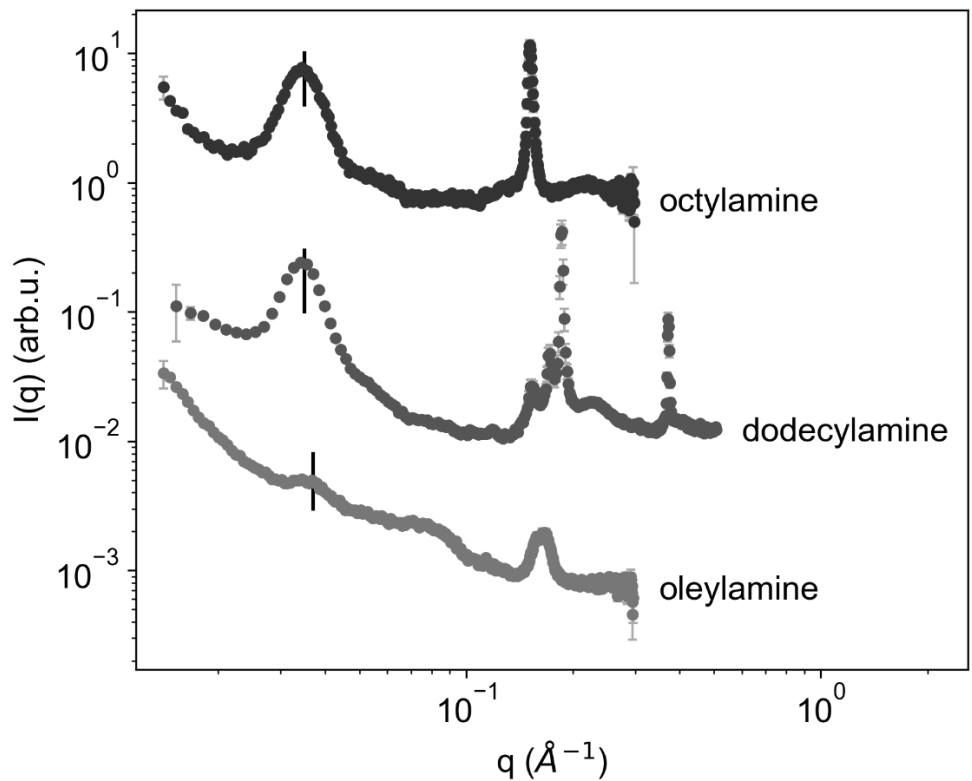

**Figure S3.** GISAXS data of perovskite NCs produced with different ligands, dried on silicon wafers. A black vertical line highlights the position of the diffraction peak due to stacking of polymer micelles.

**Table S1.** GISAXS peak positions due to stacking of polymer micelles and corresponding nearest-neighbor distances of micelles $d = \frac{4\pi}{\sqrt{3}q_{peak}}$.

| ligand | peak position (Å$^{-1}$) | d (nm) |
|---|---|---|
| octylamine | 0.035 | 20.7 |
| dodecylamine | 0.035 | 20.7 |
| oleylamine | 0.037 | 19.6 |

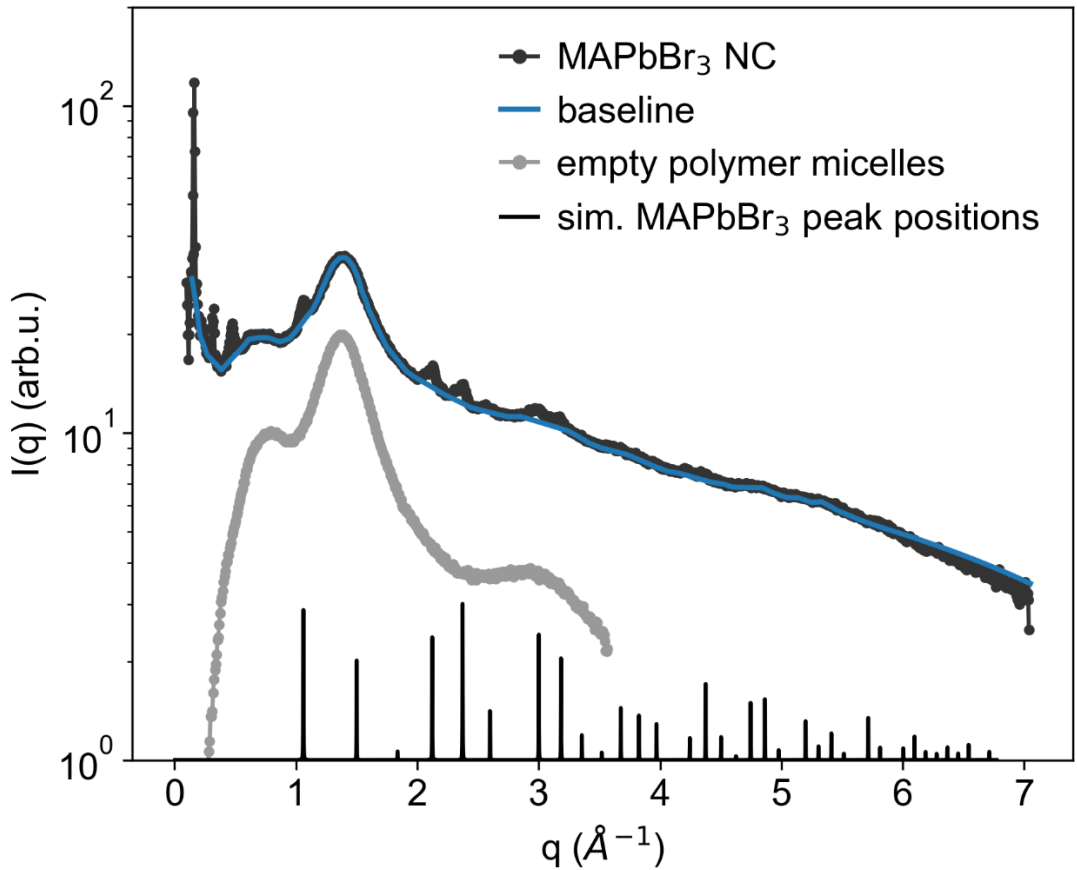

**Figure S4.** GIWAXS data of perovskite NCs (sample #3 with oleic acid and oleylamine) dried on a silicon wafer, azimuthally integrated. Blue line: baseline used for background subtraction. Gray dots: WAXS data of empty polymer micelles dried on a cyclic olefin copolymer foil shown for comparison (identical data as shown in [10]). Black lines: simulated peak positions of bulk MAPbBr3 (a = b = c = 5.92 Å, $\alpha = \beta = \gamma = 90°$).[11]

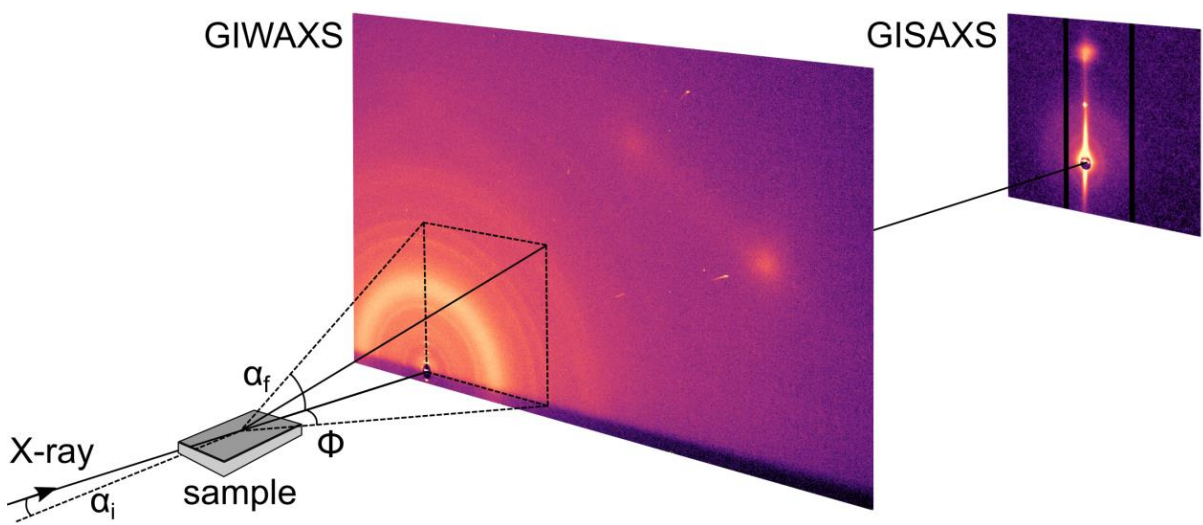

**Figure S5.** Sketch of the GIWAXS and GISAXS measurement geometries, and definition of the incident angle $\alpha_i = 0.2°$ and the exit angles $\alpha_f$ and $\Phi$, defined for every detector pixel.

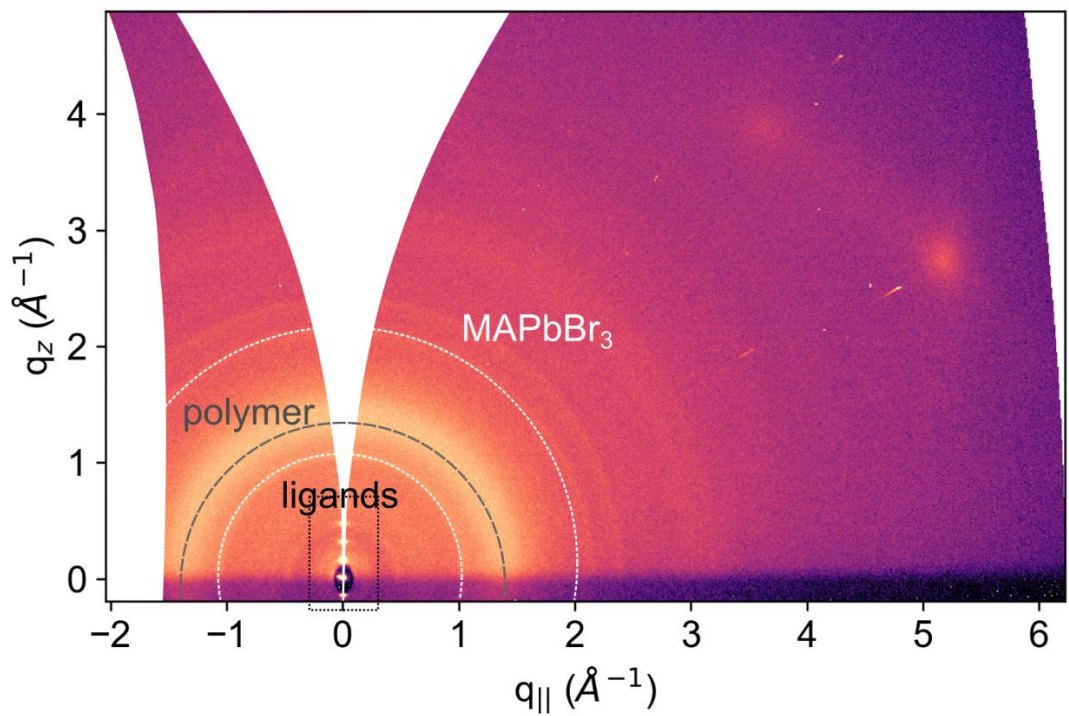

**Figure S6:** GIWAXS data of perovskite NCs (sample #3 with oleic acid and oleylamine) dried on a silicon wafer. $q_{||}$ is the scattering vector component in the substrate plane. $q_z$ is the scattering vector component perpendicular to the substrate plane. The lamellar phase of ligands is seen as regularly spaced peaks at $q_{||}$ = 0 Å$^{-1}$, $q_z$ <= 0.5 Å$^{-1}$. The polymer micelles produce a ring at q ≈ 1.4 Å$^{-1}$. The sharp rings between 1 and 3.5 Å$^{-1}$ correspond to nanocrystalline MAPbBr$_3$. The two diffuse peaks at $q_{||}$ = 3.7 Å$^{-1}$, $q_z$ = 4 Å$^{-1}$ and $q_{||}$ = 5.2 Å$^{-1}$, $q_z$ = 2.6 Å$^{-1}$ come from thermal diffuse scattering of the silicon substrate.

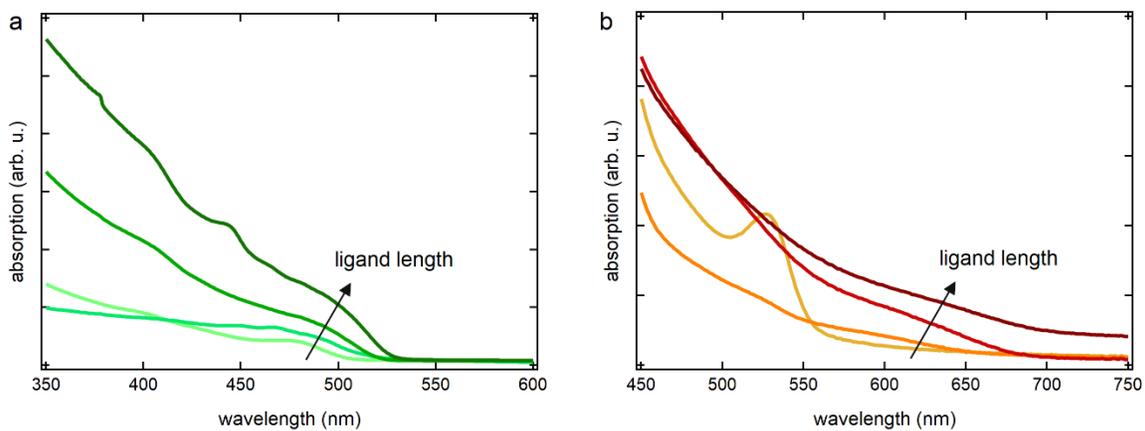

**Figure S7.** UV-Vis spectra of MAPbBr$_3$ NC dispersions (a) and MAPbI$_3$ NC dispersions (b) with different ligand lengths.

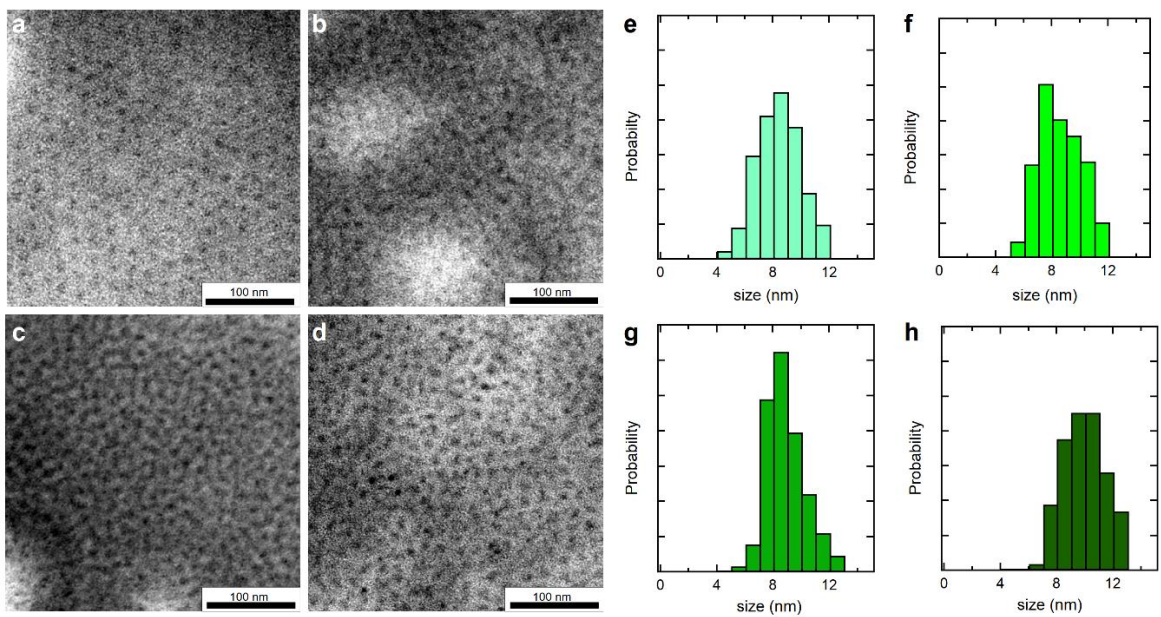

**Figure S8.** TEM images and corresponding size distributions of encapsulated MAPbBr$_3$ NCs synthesized with different ligand lengths (a, e: hexylamine, b, f: octylamine, c, g: dodecylamine, d, h: oleylamine).

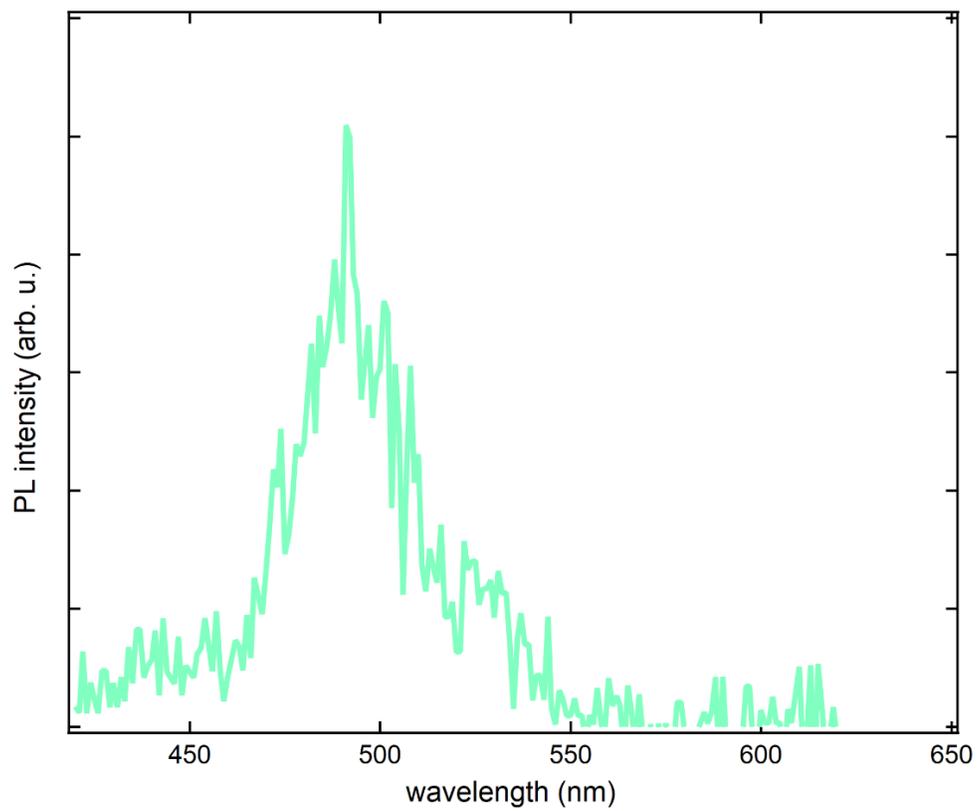

**Figure S9.** PL spectrum of encapsulated MAPbBr$_3$ NCs synthesized with butylamine and oleic acid.

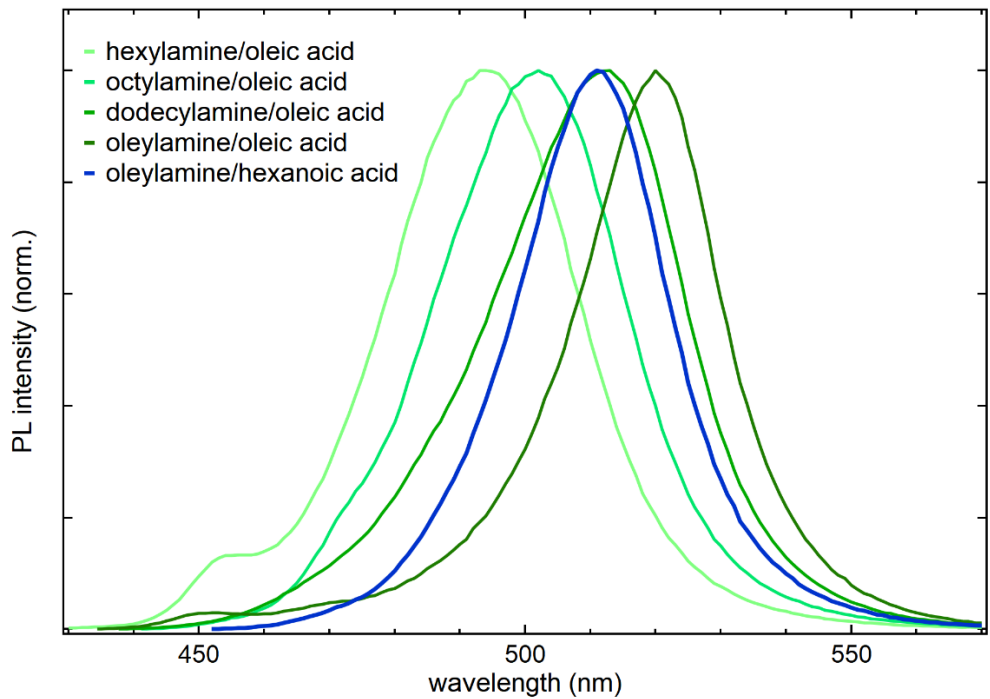

**Figure S10.** PL spectra of MAPbBr$_3$ NCs with different ligands. Hexanoic acid was used instead of oleic acid (blue line).

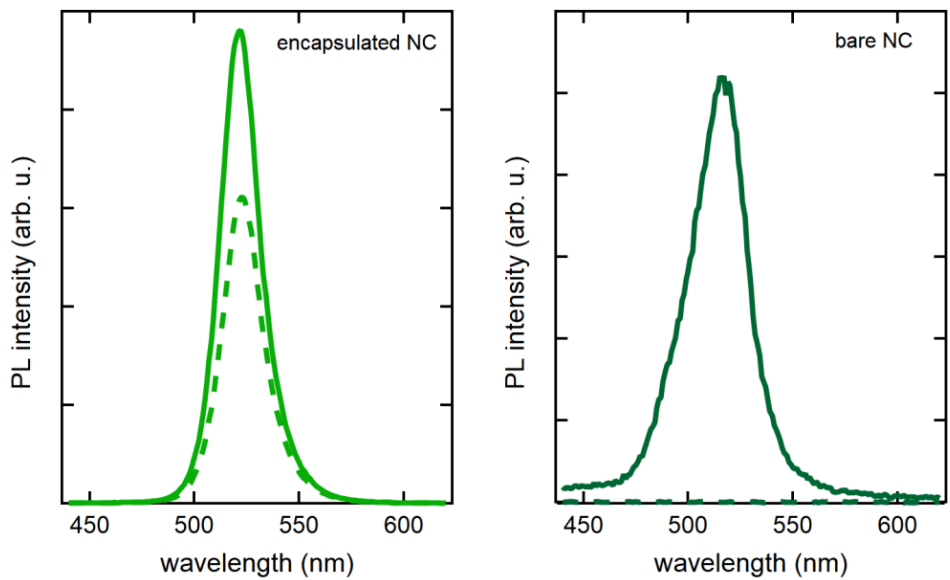

**Figure 11**. Left: PL stability of doubly protected perovskite NCs before (solid line) and five hours after (dashed line) heating to 100°C on a hot plate. Right: PL stability of bare NCs before (solid line) and one hour after (dashed line) heating to 100°C on a hot plate.

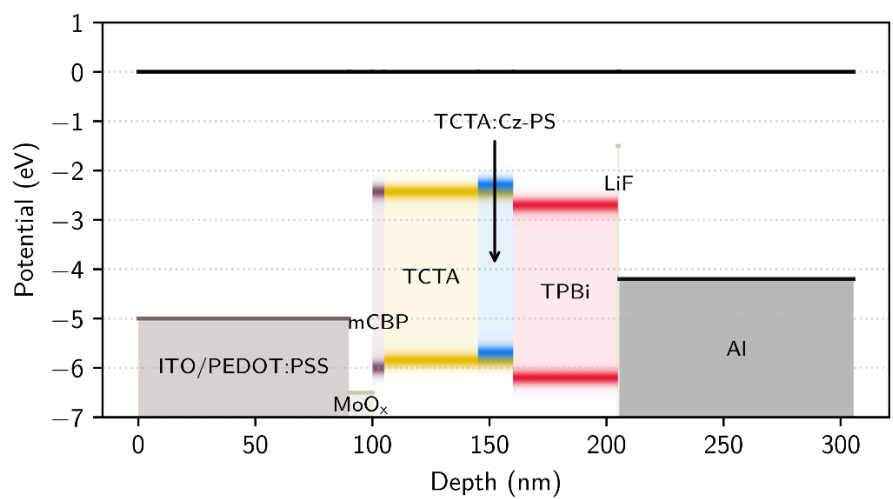

**Figure S12**. Energy level diagram and layer thicknesses of the used deep-blue OLED.

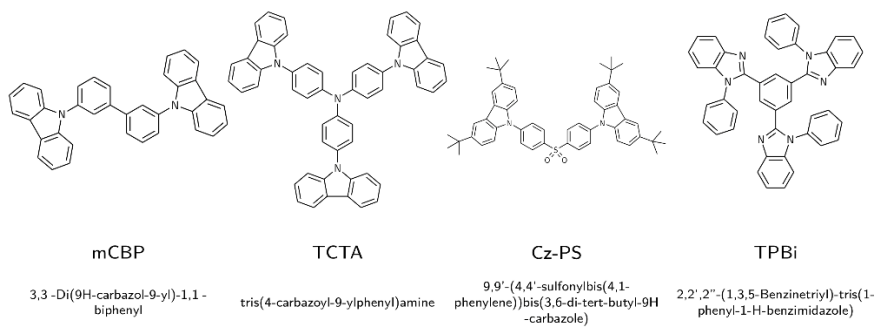

**Figure S13.** Full names and Lewis structures of the used organic semiconductors.

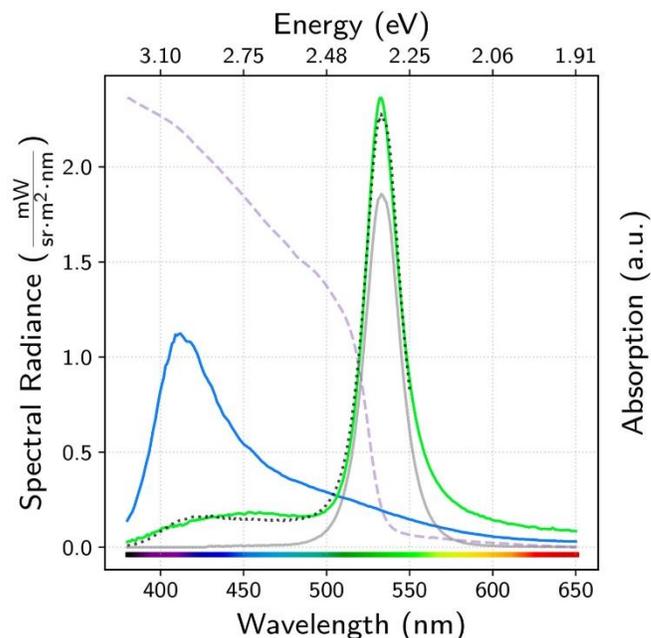

**Figure S14**. Electroluminescence (EL) spectra of the pristine blue OLED (blue) as well as the green downconverter OLED (green) measured at an applied current density of 25mA/cm². A superposition (black dotted line) of the perovskite micelles' photoluminescence and the initial OLEDs' EL spectrum weighted with the micelles' absorption dashed grey line) yields an excellent fit to the experimental DC-OLED spectrum.

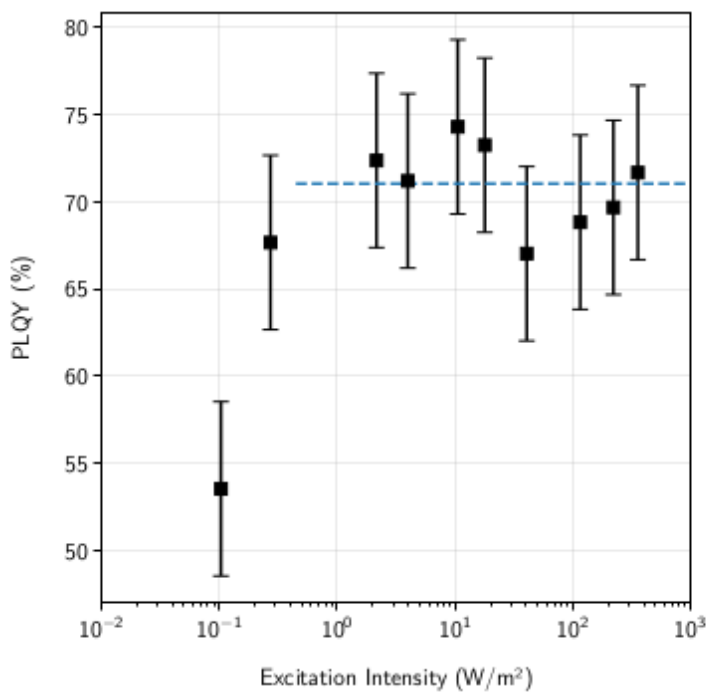

**Figure S15**. PLQY of a neat perovskite micelles DC film on a quartz-glass substrate depending on the excitation intensity of a 442 nm laser source. The average PLQY at laser fluences higher than 1 W/m² amounts to about 71%.